# Potential Theory for Directed Networks

Qian-Ming Zhang[1], Linyuan Lü[2,3]*, Wen-Qiang Wang[1], Yu-Xiao Zhu[1], Tao Zhou[1]

[1] Web Sciences Center, School of Computer Science and Engineering, University of Electronic Science and Technology of China, Chengdu, People's Republic of China,
[2] Institute of Information Economy, Alibaba Business College, Hangzhou Normal University, Hangzhou, People's Republic of China, [3] Department of Physics, University of Fribourg, Chemin du Musée 3, Fribourg, Switzerland

**Abstract**

Uncovering factors underlying the network formation is a long-standing challenge for data mining and network analysis. In particular, the microscopic organizing principles of directed networks are less understood than those of undirected networks. This article proposes a hypothesis named *potential theory*, which assumes that every directed link corresponds to a decrease of a unit potential and subgraphs with definable potential values for all nodes are preferred. Combining the potential theory with the clustering and homophily mechanisms, it is deduced that the Bi-fan structure consisting of 4 nodes and 4 directed links is the most favored local structure in directed networks. Our hypothesis receives strongly positive supports from extensive experiments on 15 directed networks drawn from disparate fields, as indicated by the most accurate and robust performance of Bi-fan predictor within the link prediction framework. In summary, our main contribution is twofold: (i) We propose a new mechanism for the local organization of directed networks; (ii) We design the corresponding link prediction algorithm, which can not only testify our hypothesis, but also find out direct applications in missing link prediction and friendship recommendation.

**Funding:** This work is partially supported by the National Natural Science Foundation of China under grant numbers 11075031 and 11205042, and the EU FET-Open project QLectives under grant number 231200. The funders had no role in study design, data collection and analysis, decision to publish, or preparation of the manuscript.

**Competing Interests:** The authors have declared that no competing interests exist.

* E-mail: linyuan.lue@unifr.ch

## Introduction

Many social, biological and technological systems can be well described by networks, where nodes represent individuals and links denote the relations or interactions between nodes. The study of structure and functions of networks has therefore become a common focus of many branches of science [1]. A big challenge attracting increasing attention in the recent decade is to uncover the mechanisms underlying the formation of networks [2]. Macroscopic mechanisms include the rich-get-richer [3], the good-get-richer [4], the stability constrains [5], and so on, while microscopic mechanisms include homophily [6], clustering [7], balance theory [8], and so on. Mechanisms can also play a part in regulating the mesoscopic structure, like the formation and transformation of groups and communities [9–11]. Real networks usually result from a hybrid of several mechanisms, for example, new nodes may form links according to the rich-get-richer mechanism, and simultaneously, new links among old nodes could be a consequence of the mechanism of clustering [12].

The so called clustering mechanism declares that two nodes have a high probability of making a link between them if they share some common neighbors [13]. This mechanism is indirectly supported by increasing evidences of high clustering coefficients (the clustering coefficient of a node is defined as the density of links among its neighbors, and the clustering coefficient of the network is the average of all nodes' clustering coefficients [14]) of disparate networks [7]. Through investigation on a social network consisting of 43,553 university members, Kossinets and Watts [15] found direct evidence that two students sharing more common acquaintances are more likely to become acquaintance with each other.

The clustering mechanism also works for directed networks, for example, in Twitter, more than 90% of new links are added between nodes sharing at least one common neighbor [16]. In addition, evolving network models driven by common neighbors could reproduce some significant features of both directed and undirected networks [17,18].

Homophily mechanism states the observed tendency of people to communicate with others of similar profiles or experiences [6]. Experiments on social networks strongly support this mechanism. Positive evidences come from various examples, such as an acquaintance network of university members [15], a large-scale instant-messaging network containing $1.8 \times 10^8$ individuals [19], friendship networks of a set of American high schools [20], a social network of a cohort of college students in Facebook [21], and so on. A variety of characteristics, such as race, tastes for music and movies, grade, age, location, language and sharing experience, are significant to the link formation. Homophily mechanism also plays a role in other kinds of networks, for example, in directed document networks, links (e.g., hyperlinks between web pages and citations between articles) tend to connect similar documents in content [22]. In some literature, the clustering mechanism is considered as a special case of homophily mechanism, where two nodes having some common neighbors are recognized as being in similar network surroundings. In this article, we prefer to distinguish these two mechanisms. Recent experiments on directed social networks show that the clustering mechanism may be even stronger than the homophily mechanism [23].

Reciprocity mechanism is the tendency of nodes to response to incoming links by creating links to the source [24]. It is a specific mechanism for some directed networks, but not applicable everywhere. For example, the reciprocity mechanism plays a significant role in the growth of social networks of Facebook-like community [25] and Flickr [26], but it has much less impacts on Slashdot [27] and it does not work at all on food webs [28].

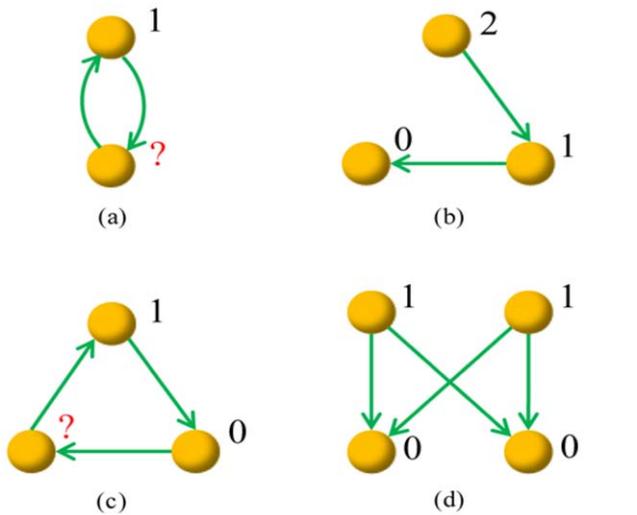

**Figure 1. Illustration of four example graphs.** Graphs (b) and (d) are potential-definable, and the numbers labeled beside nodes are example potentials. Graphs (a) and (c) are not potential-definable, and if we set the top nodes' potential to be 1, some nodes' potentials cannot be determined according to the constrain that a directed link is always associated with a decrease of a unit potential.
doi:10.1371/journal.pone.0055437.g001

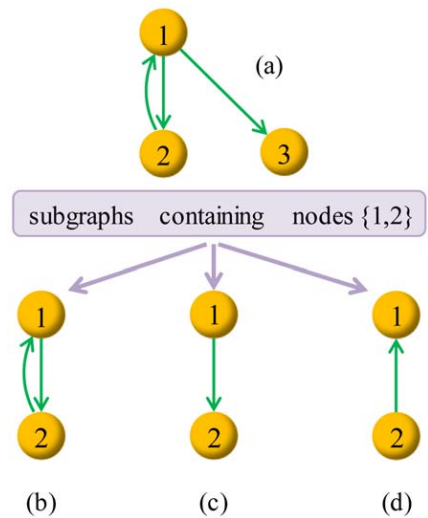

**Figure 2. Considering subgraphs of (a) that contains nodes {1,2}.** If we only consider the deduced subgraph, (b) is the unique one, while in our method, graphs (b), (c) and (d) are all subgraphs under consideration. Notice that, the empty graph containing nodes 1 and 2 and no link is also a subgraph of (a) according to our definition.
doi:10.1371/journal.pone.0055437.g002

This article focuses on directed networks. Examples of directed networks are numerous: the world wide web is made up of directed hyperlinks, the food webs consist of directed links from predators to preys, and in the microblogging social networks, fans form links pointing to their opinion leaders. High reciprocity is a specific property for some directed networks, in addition, the formation of directed links also obey the aforementioned mechanisms, for example, users in Twitter are likely to form links to neighbors of their neighbors and to friends of their friends in near ages, which are in accordance with the clustering and homophily mechanisms [16]. Besides a few representative works on local organizations (e.g., loops, small-order subgraphs, etc.) of directed networks [29–33], link formation of directed networks receives less attention and has not been well understood compared with undirected networks. Here we propose a hypothesis of link formation for general directed networks, named *potential theory*. Combining the potential theory with the clustering and homophily mechanisms, we could deduce a certain preferred subgraph. We apply the link prediction approach [34] to verify our deduction. That is, we hide a fraction of links and predict them by assuming that a link generating more preferred subgraphs is of a higher probability to exist (see details in **Methods and Materials**). Experiments on disparate directed networks ranging from large-scale social networks containing millions of individuals to small-scale food webs consisting of a hundred of species show that the prediction according to the preferred subgraph is more accurate and robust than prediction according to other comparable subgraphs. Besides the insights of the underlying mechanism for directed network formation, our work could find applications in friendship recommendation for social networks and missing link prediction for biological networks.

## Results

### Potential Theory

A graph is called potential-definable if each node can be assigned a potential such that for every pair of nodes $i$ and $j$, if there is a link from $i$ to $j$, then $i$'s potential is a unit higher than $j$. Clearly, a link is potential-definable yet a graph containing reciprocal links is not potential-definable. Figure 1 illustrates some example graphs with orders from 2 to 4, where graphs (a) and (c) are not potential-definable and graphs (b) and (d) are potential-definable. Notice that, the condition "potential-definable" is only meaningful for a very small graph since a graph consisting of many nodes is very probably not potential-definable. Although potential-definable networks are always acyclic, the directed acyclic networks [35] are usually not potential definable. For example, the feed forward loops are directed acyclic networks but not potential-definable.

The potential theory claims that a link that can generate more potential-definable subgraphs is more significant and thus of a higher probability to appear. Our definition of subgraph is more general than the traditional one. Given a directed graph $\mathbb{D}(V,E)$ with $V$ and $E$ the sets of nodes and directed links. A graph $\mathbb{D}'(V',E')$ is called a deduced subgraph of $\mathbb{D}$ if $V' \subset V$ and $E'$ contains all the links in $E$ that connect two nodes in $V'$. Our definition only requires $V' \subset V$ and $E' \subset E$, that is, $E'$ is not necessary to include all links connecting nodes in $V'$. As shown in figure 2, (b), (c) and (d) are subgraphs of (a) according to our definition, but only (b) is a deduced subgraph of (a).

Since any graph containing reciprocal links is not potential-definable, here we do not take into account the reciprocity mechanism. The clustering mechanism prefers short loops (not necessary to be directed loops) and it only works for local surrounding, and thus we only consider loop-embedded subgraphs with orders 3 and 4. Two nodes connected by reciprocal links are not treated as loops. To avoid the repeated count, we only consider the minimal loop-embedded subgraphs that do not contain loop-embedded subgraphs themselves.

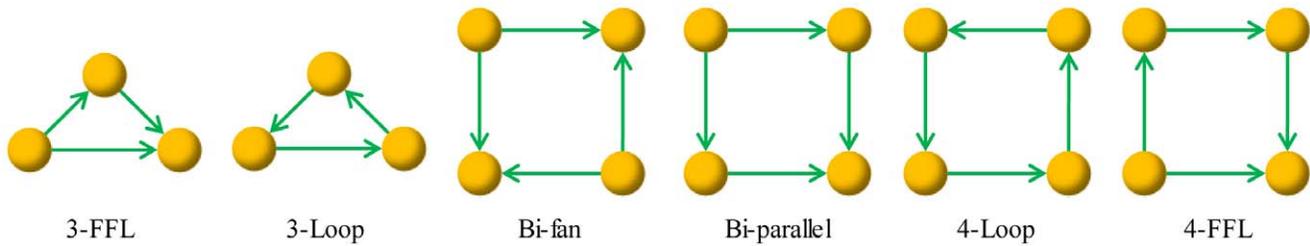

**Figure 3. All the six minimal loop-embedded subgraphs of orders 3 and 4.** They are named after Ref. [29], where 3-FFL and 4-FFL stand for three-order and four-order feed forward loops, and 3-Loop and 4-Loop mean three-order and four-order feedback loops, respectively.
doi:10.1371/journal.pone.0055437.g003

Figure 3 illustrates all the six different minimal loop-embedded subgraphs of orders 3 and 4. These subgraphs are named after Ref. [29] but our motivation is different from motif analysis and we adopt a different definition of subgraph (In Ref. [29] they only consider deduced subgraph). Among these six subgraphs, only Bi-fan and Bi-parallel are potential-definable. Since generally we could not obtain the explicit attributes of nodes, the homophily mechanism here only refers to the homogeneity in topology related to the potential levels. In a potential-definable subgraph, two nodes with the same potential cannot directly connect to each other and thus the homophily mechanism only works when we consider each subgraph as a whole. Specifically, a subgraph is more homogeneous if the nodes therein are of fewer potential levels. For Bi-fan the links are equivalent to each other and nodes are of two different potentials, while in Bi-parallel, links are different (two are from high-potential nodes to moderate-potential nodes, and the other two are from moderate-potential nodes to low-potential nodes) and nodes are of three different potentials. According to the assigned potentials, we could say the Bi-fan structure is more homogeneous (of fewer potential levels) than the Bi-parallel structure, then the homophily mechanism prefers the former one.

In a word, taking into account the potential theory, together with the clustering and homophily mechanisms, it is thought that the Bi-fan subgraph is the most preferred one and a link that can generate more Bi-fan subgraphs should be of higher probability to exist. This hypothesis receives strongly positive supports as indicated by the most accurate and robust performance of Bi-fan predictor within the link prediction framework. Figure 4 illustrates the selecting procedure for the final winner Bi-fan, as well as the respective contributions of the three mechanisms.

### Experimental Results

Corresponding to these six subgraphs we get 12 individual predictors by removing one link from every subgraph (S1–S12, see figure 5). To evaluate the accuracy of a predictor, a network is divided into two parts – training set and testing set. Denote one pair of disconnected nodes in the network as a nonexistent link, then all links can be classified into three categories: observed links are the ones in the training set, missing links are the ones in the testing set, and nonexisting links are the remain links. All the missing links and nonexisting links constitute the set of non-observed links. A good predictor will assign higher scores to missing links than nonexistent ones. We adopt the Area under the Receiver operating characteristic Curve (AUC) to evaluate the prediction accuracy: a higher AUC value corresponds to a better predictor. Please see details about the link prediction algorithm and the evaluation metric for algorithmic performance in **Methods and Materials**.

Table 1 shows the prediction accuracy, measured by AUC values, of all the 12 individual predictors. In 14 out of 15 real networks, except Youtube, the predictor $S_5$ performs best. The advantage of the predictor $S_5$ to others is usually remarkable, while for Youtube, the performance of $S_5$ is very close to the optimal one, $S_{12}$. The last row of Table 1 shows the average AUC values, which again emphasizes the great advantage of $S_5$. Roughly speaking, the very simple rule – a link generating more Bi-fan subgraphs has higher probability to exist – is nearly 90% right.

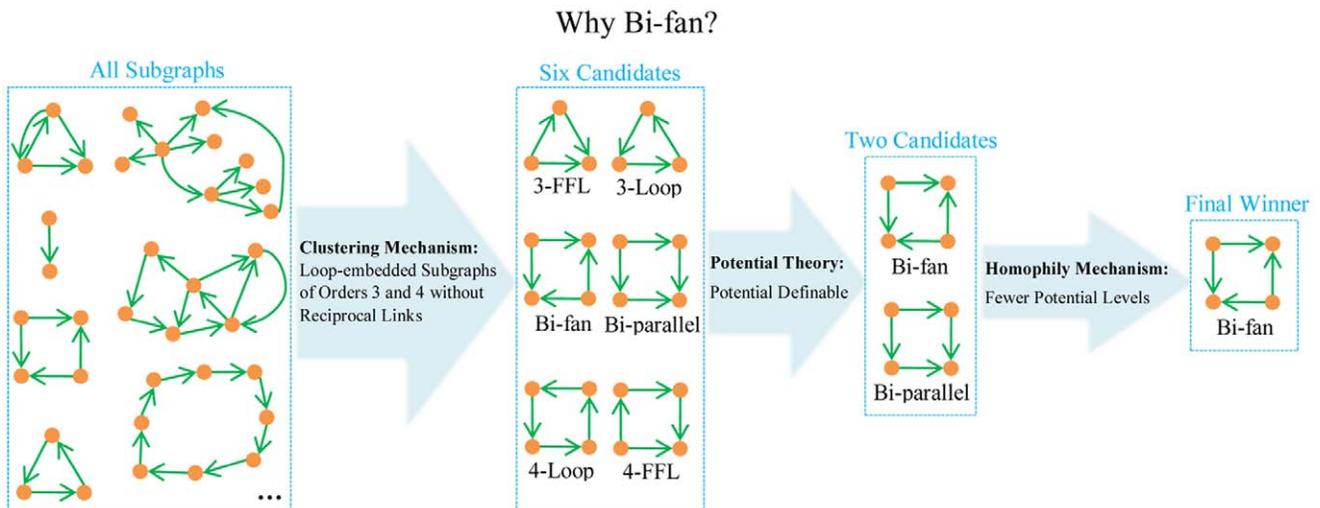

**Figure 4. Illustration of the reason why Bi-fan is selected to be the final winner according to the homophily mechanism, clustering mechanism and potential theory.**
doi:10.1371/journal.pone.0055437.g004

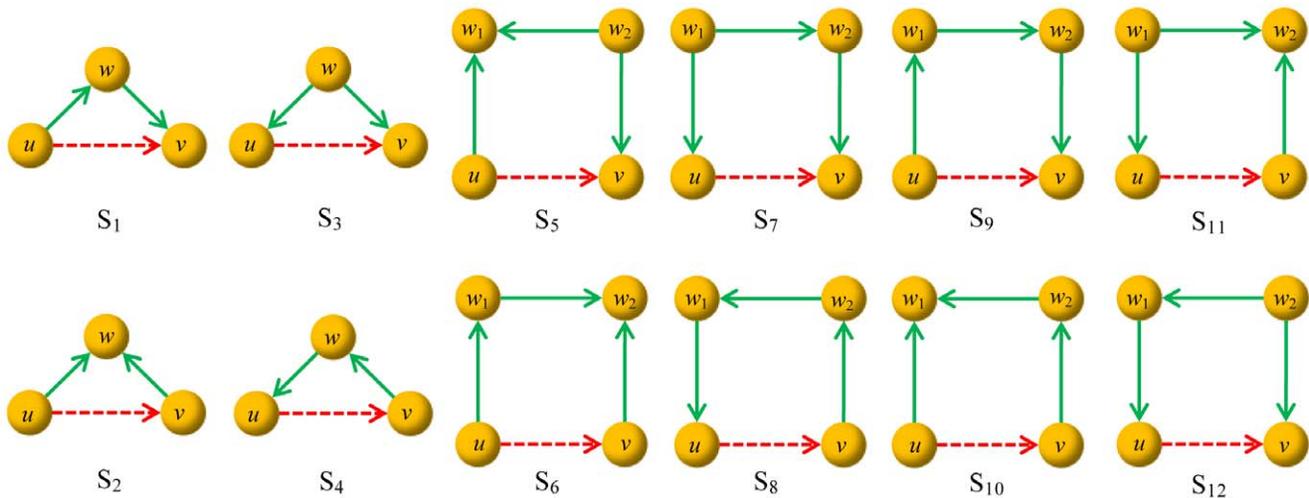

**Figure 5. Illustration of the twelve predictors corresponding to the subgraphs shown in figure 3.** The red dashed arrows represent the links removed from the original subgraphs. The relations are as follows: {$S_1$, $S_2$, $S_3$} ⇔ 3-FFL, {$S_4$} ⇔ 3-Loop, {$S_5$} ⇔ Bi-fan, {$S_6$, $S_7$} ⇔ Bi-parallel, {$S_8$} ⇔ 4-Loop, {$S_9$, $S_{10}$, $S_{11}$, $S_{12}$} ⇔ 4-FFL.
doi:10.1371/journal.pone.0055437.g005

Table 2 shows the comparison of the prediction accuracy of some hybrid predictors. We explain again that the predictor $S_1 + S_2 + S_3$ means that the score of a non-observed link is defined as the number of created $S_1$, $S_2$ and $S_3$ resulting from the addition of this link. In fact, the six predictors in Table 1 correspond to the six minimal loop-embedded subgraphs in figure 3. Therefore, Table 1 directly gives the comparison of the six candidate subgraphs. Again, Bi-fan wins.

Looking at the results presented in Table 1 and Table 2, another significant advantage of the Bi-fan structure is the high robustness, that is to say, even when the predictor $S_5$ is not the best in some cases, its performance is very close to the optimal one. In contrast, for any other predictor, no matter what predictor–an individual predictor or a hybrid one, it is very sensitive to the network structure, and will occasionally give very bad predictions.

**Table 1.** AUC values of the 12 predictors shown in figure 5.

| Datasets | $S_1$ | $S_2$ | $S_3$ | $S_4$ | $S_5$ | $S_6$ | $S_7$ | $S_8$ | $S_9$ | $S_{10}$ | $S_{11}$ | $S_{12}$ |
|---|---|---|---|---|---|---|---|---|---|---|---|---|
| FW1 | 0.7400 | 0.4634 | 0.6156 | 0.4903 | **0.9066** | 0.6147 | 0.7811 | 0.4172 | 0.7848 | 0.4254 | 0.3236 | 0.5697 |
| FW2 | 0.7629 | 0.5507 | 0.6367 | 0.4809 | **0.8964** | 0.6965 | 0.7838 | 0.4972 | 0.6822 | 0.4255 | 0.3818 | 0.5456 |
| FW3 | 0.7333 | 0.5364 | 0.5675 | 0.3997 | **0.9105** | 0.7282 | 0.7757 | 0.4303 | 0.6683 | 0.3517 | 0.3210 | 0.4532 |
| C.elegans | 0.7886 | 0.7127 | 0.7569 | 0.5671 | **0.8679** | 0.7686 | 0.7991 | 0.5755 | 0.7990 | 0.6528 | 0.6667 | 0.7591 |
| SmaGri | 0.7074 | 0.6517 | 0.6905 | 0.4922 | **0.8852** | 0.7108 | 0.7476 | 0.4851 | 0.6677 | 0.6242 | 0.5982 | 0.5761 |
| Kohonen | 0.6693 | 0.6124 | 0.6642 | 0.4991 | **0.8605** | 0.6333 | 0.7335 | 0.4985 | 0.6148 | 0.5614 | 0.5778 | 0.5946 |
| SciMet | 0.6462 | 0.6192 | 0.6371 | 0.4980 | **0.8371** | 0.6672 | 0.7045 | 0.4968 | 0.5977 | 0.5794 | 0.5753 | 0.5895 |
| PB | 0.9025 | 0.8181 | 0.8243 | 0.6948 | **0.9595** | 0.8659 | 0.8679 | 0.7518 | 0.9479 | 0.8349 | 0.7616 | 0.8584 |
| Delicious | 0.7298 | 0.7077 | 0.7192 | 0.6577 | **0.7839** | 0.7141 | 0.7344 | 0.6739 | 0.7378 | 0.7081 | 0.7046 | 0.7273 |
| Youtube | 0.7518 | 0.7453 | 0.7522 | 0.7456 | 0.8517 | 0.8422 | 0.8576 | 0.8442 | 0.8505 | 0.8430 | 0.8507 | **0.8624** |
| FriendFeed | 0.8801 | 0.7503 | 0.7382 | 0.5895 | **0.9766** | 0.7863 | 0.8100 | 0.7150 | 0.9690 | 0.8324 | 0.7318 | 0.8027 |
| Epinions | 0.8273 | 0.8326 | 0.8081 | 0.7460 | **0.9101** | 0.8969 | 0.8843 | 0.8584 | 0.8995 | 0.8956 | 0.8804 | 0.8831 |
| Slashdot | 0.7164 | 0.7133 | 0.7124 | 0.7072 | **0.9035** | 0.8984 | 0.8982 | 0.8925 | 0.9009 | 0.8982 | 0.8926 | 0.8985 |
| Wikivote | 0.9073 | 0.7448 | 0.7470 | 0.5962 | **0.9699** | 0.7679 | 0.7451 | 0.6209 | 0.9583 | 0.7562 | 0.6096 | 0.7468 |
| Twitter | 0.8937 | 0.7226 | 0.8289 | 0.7586 | **0.9734** | 0.7856 | 0.9444 | 0.7545 | 0.9582 | 0.8108 | 0.7557 | 0.9527 |
| Average | 0.7771 | 0.6787 | 0.7133 | 0.5949 | **0.8995** | 0.7584 | 0.8045 | 0.6341 | 0.8024 | 0.6800 | 0.6421 | 0.7213 |

The best performance for each network is emphasized in bold. Each number is obtained by averaging over 50 implementations with independently random partitions of training set and testing set.
doi:10.1371/journal.pone.0055437.t001

Table 2. AUC values of the six subgraphs shown in figure 3.

| Datasets | $S_1+S_2+S_3$ | $S_4$ | $S_5$ | $S_6+S_7$ | $S_8$ | $S_9+S_{10}+S_{11}+S_{12}$ |
|---|---|---|---|---|---|---|
| FW1 | 0.6953 | 0.4903 | **0.9066** | 0.8462 | 0.4172 | 0.4653 |
| FW2 | 0.7241 | 0.4809 | **0.8964** | 0.8490 | 0.4972 | 0.4674 |
| FW3 | 0.6649 | 0.3997 | **0.9105** | 0.8586 | 0.4303 | 0.3283 |
| C.elegans | 0.8666 | 0.5671 | **0.8679** | 0.8403 | 0.5755 | 0.7736 |
| SmaGri | 0.8400 | 0.4922 | **0.8852** | 0.8154 | 0.4851 | 0.7291 |
| Kohonen | 0.8091 | 0.4991 | **0.8605** | 0.7779 | 0.4985 | 0.7039 |
| SciMet | 0.7874 | 0.4980 | **0.8371** | 0.7872 | 0.4968 | 0.7187 |
| PB | 0.9275 | 0.6948 | **0.9595** | 0.9029 | 0.7518 | 0.9122 |
| Delicious | 0.7621 | 0.6577 | 0.7839 | 0.7743 | 0.6739 | **0.7893** |
| Youtube | 0.7526 | 0.7456 | 0.8517 | 0.8593 | 0.8442 | **0.8625** |
| FriendFeed | 0.7937 | 0.5895 | **0.9766** | 0.9151 | 0.7150 | 0.9240 |
| Epinions | 0.8682 | 0.7460 | 0.9101 | 0.9131 | 0.8584 | **0.9174** |
| Slashdot | 0.7422 | 0.7072 | 0.9035 | 0.9048 | 0.8925 | **0.9083** |
| Wikivote | 0.9330 | 0.5962 | **0.9699** | 0.8607 | 0.6209 | 0.9288 |
| Twitter | 0.8251 | 0.7586 | **0.9734** | 0.9351 | 0.7545 | 0.9484 |
| Average | 0.7995 | 0.5949 | **0.8995** | 0.8560 | 0.6341 | 0.7585 |

The best performance for each network is emphasized in bold. Each number is obtained by averaging over 50 implementations with independently random partitions of training set and testing set.
doi:10.1371/journal.pone.0055437.t002

## Discussion

This article studied the underlying mechanism of the link formation for directed networks. We presented a hypothesis named potential theory, which claims that a link that can generate more potential-definable subgraphs is of a higher probability to appear. This mechanism cannot be solely used to infer network structure for there are too many potential-definable subgraphs (e.g., directed paths of any lengths are potential definable). Therefore, we also take into account two well-known local mechanisms: clustering and homophily. By combining the three mechanisms, it is inferred that Bi-fan is the most preferred subgraph in directed networks. Via comparison of the link prediction accuracies of 12 individual predictors as well as six minimal loop-embedded subgraphs, Bi-fan performs best: not only for its higher AUC value than others, but also for its robustness, namely for disparate testing networks, its performance is either the best or very close to the best. Notice that though the experimental results provided supportive evidences, they can only be considered as a necessary condition, but not a sufficient condition or a solid proof for the potential theory.

The local driven mechanisms underlying directed network formation are less understood compared with those for undirected networks. This kind of study is thus of theoretical significance, and our work provided insights into the microscopic architecture of directed networks. Although the potential theory is more complicated than the clustering and homophily mechanisms as well as the balance theory, its meaning is easy to be captured, that is, the potential-definable property implies a local hierarchy and the potential value of a node indicates its level in the hierarchical structure. For example, the directed loops are not hierarchy-embedded and the directed path is strictly hierarchically organized; the former is not potential-definable and the later is potential-definable. The hierarchical organization is a well-known macroscopic feature for many undirected [36,37] and directed [38,39] networks, and our work indicates that for directed networks, nodes tend to be locally self-organized in a hierarchical manner. We guess this kind of microscopic hierarchical organization will contribute to the macroscopic hierarchical structure. In the near future, we will study more data sets in a more detailed way to check whether the potential theory and our hypothesis about hierarchical organization are valid or not and to see the applicable range (to which networks it works and to what extent it can explain the network formation) of the potential theory.

Lastly, we would like to say again that the link prediction problem is very fundamental to both information filtering and network analysis [34,40], and it could find out countless applications. In this work, we applied the link prediction approach to evaluate driven mechanisms of network formation, at the same time, our method can be directly applied to predicting missing links and recommending friendships for large-scale directed

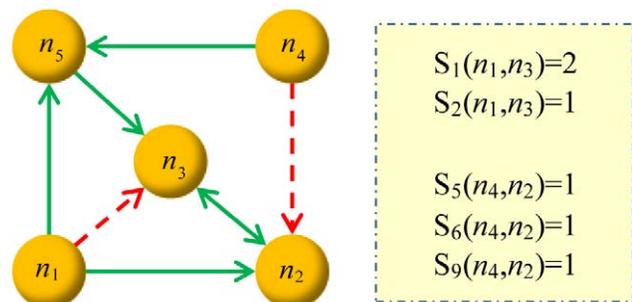

Figure 6. Illustration of the scores of links according to our method. The red dashed arrows are probe links. If we adopt the predictor $S_1$, the scores for $n_1 \to n_3$ and $n_4 \to n_2$ are $S_1(n_1 \to n_3) = 2$ ($n_1 \to n_5 \to n_3$ and $n_1 \to n_2 \to n_3$) and $S_1(n_4 \to n_2) = 0$, respectively. More examples are as follows: $S_2(n_1 \to n_3) \blacktriangleright \{n_1 \to n_2 \leftarrow n_3\}$; $S_5(n_4 \to n_2) \blacktriangleright \{n_4 \to n_5 \leftarrow n_1 \to n_2\}$; $S_6(n_4 \to n_2) \blacktriangleright \{n_4 \to n_5 \to n_3 \leftarrow n_2\}$; $S_9(n_4 \to n_2) \blacktriangleright \{n_4 \to n_5 \to n_3 \to n_2\}$.
doi:10.1371/journal.pone.0055437.g006

**Table 3.** The basic structural features of the studied 15 real networks.

| Networks | $|V|$ | $|E|$ | $k_{max}^{in}$ | $k_{max}^{out}$ | $\langle k \rangle$ | $\langle d \rangle$ | $C$ | References |
|---|---|---|---|---|---|---|---|---|
| FW1 | 69 | 916 | 63 | 44 | 13.3 | 2.84 | 0.552 | [42] |
| FW2 | 97 | 1492 | 90 | 46 | 15.4 | 2.86 | 0.468 | [43] |
| FW3 | 128 | 2137 | 110 | 63 | 16.7 | 2.90 | 0.335 | [44] |
| C.elegans | 297 | 2345 | 134 | 39 | 7.9 | 3.85 | 0.292 | [45] |
| SmaGri | 1024 | 4919 | 89 | 232 | 4.8 | 4.61 | 0.302 | [46] |
| Kohonen | 3704 | 12683 | 51 | 735 | 3.4 | 5.64 | 0.252 | [46] |
| SciMet | 2678 | 10381 | 121 | 104 | 3.9 | 6.40 | 0.174 | [46] |
| PB | 1222 | 19021 | 337 | 256 | 15.6 | 4.08 | 0.320 | [47] |
| Delicious | 571686 | 1668233 | 2767 | 11168 | 2.9 | 8.65 | 0.202 | [48] |
| Youtube | 1134890 | 4942035 | 25519 | 28644 | 4.4 | 7.17 | 0.081 | [49] |
| FriendFeed | 512889 | 19810241 | 31045 | 96659 | 38.6 | 4.92 | 0.215 | [50] |
| Epinions | 75877 | 508836 | 3035 | 1801 | 6.7 | 6.45 | 0.138 | [51] |
| Slashdot | 77360 | 828161 | 2539 | 2507 | 10.7 | 5.62 | 0.056 | [52] |
| Wikivote | 7066 | 103663 | 457 | 893 | 14.7 | 4.77 | 0.142 | [53,54] |
| Twitter | 11241 | 732193 | 5665 | 3633 | 65.14 | 2.7 | 0.162 | [55] |

$|V|$ and $|E|$ are the number of nodes and links, $k_{max}^{in}$ and $k_{max}^{out}$ are the maximum of in-degree and out-degree of all nodes, and $\langle k \rangle$ is the average degree of all nodes (average in-degree equals average out-degree). $\langle d \rangle$ and $C$ are the 90-percentile effective diameter [56] and the clustering coefficient for directed networks [57].
doi:10.1371/journal.pone.0055437.t003

networks, since the accuracy of our method is much higher than the common-neighbor-based methods as indicated by the performance of predictors $S_1$, $S_2$, $S_3$ and $S_4$.

## Materials and Methods

### Link Prediction Algorithm

Given a directed network $\mathbb{D}(V,E)$, the fundamental task of a link prediction algorithm is to give a rank of all non-observed links in the set $U \setminus E$, where $U$ is the universal set containing all $|V|(|V|-1)$ possible directed links. If one wants to find out missing links or recommend friendships, one can go for the links with the highest ranks. The mainstream method is to assign each non-observed link a score, and the one with higher score ranks ahead.

We design the predictors corresponding to the six minimal loop-embedded subgraphs shown in figure 3. By removing one link from every subgraph, we get twelve predictors as shown in figure 5. If we adopt the predictor $S_i$, it means the score of a non-observed link $u \rightarrow v$ is defined as the number of the $i$th subgraphs created by the addition of this link. Notice that, a link may generate ten 3-FFLs, but their roles can be different. For example, these ten 3-FFLs may include two $S_1$, three $S_2$ and five $S_3$. So if we adopt the predictor $S_2$, the score of this link is three. Therefore, if we would like to see the contribution of a link to the created 3-FFLs, we can adopt the predictor $S_1+S_2+S_3$, which means that the score of a non-observed link is defined as the total number of created $S_1$, $S_2$ and $S_3$ by this link, equivalent to the number of created 3-FFLs. Figure 6 illustrates a simple example about how we calculate the scores.

Given a predictor we can rank all the non-observed links according to their scores. To evaluate the algorithmic performance, we randomly divide the observed links $E$ into two parts: the training set $E^T$ is treated as known information while the testing set (probe set) $E^P$ is used for testing and no information therein is allowed to be used for prediction. Clearly, $E = E^T \cup E^P$ and $E^T \cap E^P = \phi$. In our experiments, the training set always contains 90% of links, and the remaining 10% of links constitute the testing set.

### Evaluation Metric

We use a standard metric, area under the receiver operating characteristic (ROC) curve [41], to test the accuracy of link prediction algorithms. It is usually abbreviated as AUC (Area Under Curve) value. This metric can be interpreted as the probability that a randomly chosen missing link (a link in $E^P$) is given a higher score than a randomly chosen nonexistent link (a link in $U \setminus E$). In the implementation, among $n$ times of independent comparisons, if there are $n'$ times the missing link having higher score and $n''$ times the missing link and nonexistent link having the same score, we define the AUC value as [34]:

$$\text{AUC} = \frac{n' + 0.5 n''}{n}.$$

If all the scores are generated from an independent and identical distribution, the AUC value should be about 0.5. Therefore, the degree to which the AUC value exceeds 0.5 indicates how much better the algorithm performs than pure chance.

### Data Description

Our experiments include 15 real directed networks drawn from disparate fields. Details are as follows and the basic structural features are presented in Table 3. If a network is unconnected, we only consider its largest weakly connected component.

**Biological networks.** Three of them are food webs, representing the predator-pray relations, and another one is a neural network of C.elegans.

- FW1 [42] – A food web consists of 69 species living in Everglades Graminoids during wet season.

- FW2 [43] – A food web consists of 97 species living in Mangrove Estuary during wet season.
- FW3 [44] – A food web consists of 128 species living in Florida Bay during dry season.
- C.elegans [45] – A neural network of the nematode worm C.elegans, in which an edge joins two neurons if they are connected by either a synapse or a gap junction.

**Information networks.** We consider networks of documents where a directed link from $i$ to $j$ means the document $i$ cites the document $j$, and a network of weblogs where a directed link stands for a hyperlink.

- Small & Griffith and Descendants (SmaGri) [46] – Citations to Small & Griffith and Descendants.
- Kohonen [46] – Articles with topic ''self-organizing maps'' or references to ''Kohonen T''.
- Scientometrics (SciMet) [46] – Articles from or citing Scientometrics.
- Political Blogs (PB) [47] – A directed network of hyperlinks between weblogs on US political blogs.

**Social networks.** All the following networks describe relationships between people.

- Delicious [48] – Delicious.com, previously known as del.i-cio.us, allows individuals to tag the bookmarks and follow other users. The studied who-follow-whom network was collected at May 2008.
- Youtube [49] – YouTube offers the greatest platform where users can share videos with others. Active users who regularly upload videos maintain a channel pages. Other users can follow those users thus forming a social network. This data was collected at January 2007.
- FriendFeed [50] – FriendFeed is an aggregator that consolidates the updates from the social media and social networking websites, social bookmarking websites, blogs and micro-blogging updates, etc. Members can manage their social networking contents with one Friend-Feed account and follow others' updates. This data set captures the who-follow-whom relationships.
- Epinions [51] – Epinions.com is a who-trust-whom online social network of a general consumer review site. Members of this site can decide whether to "trust" each other.
- Slashdot [52] – Slashdot.org is a technology-related news website known for its specific user community. This site allows individuals to tag each other as friends or foes.
- Wikivote [53,54] – Wikipedia is a free encyclopedia written collaboratively by volunteers around the world. Active users can be nominated to be administrator. A public voting begins after some users are nominated. Other users can express their positive, negative or neural idea towards all the candidates. The most voted candidate will be promoted to admin status. This process implies a social network in which users are nodes and the action of voting from someone to another demonstrates a directed link. This data is from English Wikipedia on 2794 elections.
- Twitter [55] – Twitter is an online social networking service where users can post texts within 140 characters. It also allow users to "follow" other users whereby a user can see updates from the users he follows on his twitter page. In this network, a link from user A to user B means that user A is following user B. The data used here is a sample from the whole dataset in [55].

## Acknowledgments

We acknowledge An Zeng, Changsong Zhou and Xiao-Ke Xu for helpful discussions and irradiative ideas.

## Author Contributions

Conceived and designed the experiments: QMZ LL TZ. Performed the experiments: QMZ WQW YXZ. Analyzed the data: QMZ LL WQW. Contributed reagents/materials/analysis tools: QMZ LL TZ. Wrote the paper: QMZ LL TZ.